\documentclass[12pt]{iopart}

\newtheorem{lemma}{Lemma}

\newtheorem{theorem}{Theorem}

\newcommand{\be}{\begin{equation}}
\newcommand{\ee}{\end{equation}}

\newcommand{\dif}{\mbox{\rm d}}

\begin{document}

\title[Degenerate static metrics] {A note on static
metrics: the degenerate case}

\author{Joan Josep Ferrando$^1$\
and Juan Antonio S\'aez$^2$}

\address{$^1$\ Departament d'Astronomia i Astrof\'{\i}sica, Universitat
de Val\`encia, E-46100 Burjassot, Val\`encia, Spain.}

\address{$^2$\ Departament de Matem\`atiques per a l'Economia i l'Empresa,
Universitat de Val\`encia, E-46071 Val\`encia, Spain}

\ead{joan.ferrando@uv.es; juan.a.saez@uv.es}

\begin{abstract}
We give the necessary and sufficient conditions for a 3-metric to be
the adapted spatial metric of a static vacuum solution. This work
accomplishes for the degenerate cases the already known study for the
regular ones ({\it Class. Quantum Grav.} {\bf 23} (2006) 569-571).
\end{abstract}


%
%
\vspace{1cm}
\noindent For a static metric $ g = - e^{2U(x^k)} \dif T^2 +
\gamma_{ij}(x^k) \dif x^i \dif x^j $,
the vacuum field Einstein equations reduce to the following coupled
system involving the 3-dimensional {\it spatial metric} $\gamma$ and
the {\it potential function} $U$:
\begin{equation} \label{sve}
\tr R = 0  \, , \qquad  R = \nabla \chi + \chi \otimes \chi \, ,
\qquad \chi = \dif U  \, .
\end{equation}
where $R \equiv R(\gamma)$ denotes the Ricci tensor of the metric
$\gamma$ and $\nabla$ its covariant derivative.

In \cite{bar-tod} the following question was established: what are
the necessary and sufficient conditions for a 3-metric to be the
spatial metric of a static vacuum solution? In other words, what
minimal conditions must $\gamma$ satisfy to guarantee that a $U$
exists such that the pair $\{\gamma, U\}$ is a solution of
(\ref{sve})? This problem was solved in \cite{bar-tod} for the
family of spatial metrics with a non-degenerate Ricci tensor. Here
we complete this study by solving the problem for metrics with a
degenerate Ricci tensor.

A static metric $g$ has real Weyl eigenvalues, i.e., the magnetic
part of the Weyl tensor vanishes. Moreover the electric Weyl tensor
$E \equiv E(g)$ is the Ricci tensor of the spatial metric $\gamma$,
$E = R$. Consequently:
\begin{lemma}
The spatial metric $\gamma$ has a degenerate Ricci tensor if, and
only if, the space-time metric $g$ is of Petrov-Bel type $D$.
\end{lemma}
Of course, a degenerate Ricci tensor means a double eigenvalue. We
do not consider the case of a triple eigenvalue which imposes $g$ to
be the flat metric. On the other hand, the non degenerate case
corresponds to Petrov-Bel type $I$ space-times.

The $\gamma$-Ricci identities for the vector $\chi = \dif U$ lead to
\cite{bar-tod}:
\begin{equation} \label{ricci-dU}
C_{ij} = \eta_j^{\ pq}(2 R_{ip} \gamma_{qk} + \gamma_{ip} R_{qk})
\chi^k \, ,
\end{equation}
where $C$ is the Cotton-York tensor which, under the traceless Ricci
condition, is
\begin{equation}
C_{ij} = - \eta_j^{\ pq} \nabla_q R_{ip}  \, .
\end{equation}

When the Ricci tensor is not degenerate there are three independent
equations  in the five equations (\ref{ricci-dU}). Then, the vector
$\chi$ can be obtained in terms of metric concomitants and one
arrives to the Bartnik and Tod result \cite{bar-tod}. Here, we
state it as follows:
\begin{theorem} \label{theorem-bartod}
{\bf (Bartnik and Tod)} Let $\gamma$ be a 3-dimensional metric, $R
\equiv R(\gamma)$ its Ricci tensor and $s \equiv  \tr R^2$, $t
\equiv  \tr R^3$. The necessary and sufficient conditions for
$\gamma$ to be the spatial metric of a type $I$ static vacuum
solution are:
\begin{equation} \label{bartod-1}
\tr R = 0 \, ,  \qquad \Delta \equiv 12 t^2 - 2 s^3 \not = 0  \, ,
\end{equation}
\begin{equation} \label{bartod-2}
 R =
\nabla \chi + \chi \otimes \chi \, ,  \qquad  \chi \equiv
\frac{1}{\Delta} (6 R^2 - s \gamma)^2(*[R,C]) \,
\ee $C \equiv C(\gamma)$ being the Cotton-York tensor of $\gamma$.
\end{theorem}

For a vacuum type $D$ metric the Ricci tensor of the spatial metric
$\gamma$ has a simple eigenvector $e$ and an eigenplane with
projector $\sigma$. If $\mu$ is the associated double eigenvalue, we
have:
\begin{equation} \label{ricci-D}
R = -2 \mu\, e \otimes e + \mu \sigma =  \mu(-3\, e \otimes e + \gamma) \, .
\end{equation}

If we put expression (\ref{ricci-D}) on the right-hand side of equation
(\ref{ricci-dU}) we obtain,
\begin{equation} \label{D-constraints}
C(e,e)=0\, , \qquad C_{\perp} =0 \, ,
\end{equation}
where $\perp$ denotes the $e$-orthogonal projection,
$(C_{\perp})_{ij} = \sigma_{i}^k  \sigma_{j}^l C_{kl}$.
Consequently,  there are two independent equations  in
(\ref{ricci-dU}) and we can not obtain $\chi$ from it. In fact, the
only equations that (\ref{ricci-dU}) imposes on $\chi$ can be
written as:
\begin{equation} \label{Ce}
C(e)=  3\mu *(e \wedge \chi) \, .
\end{equation}

On the other hand, from equations (\ref{D-constraints}) and the
Bianchi identities, $\nabla \cdot R =0$, we obtain:
\begin{equation} \label{nabla-e}
\nabla e = e \otimes \dif m - e(m) \gamma \,  ,  \qquad m \equiv \ln
|\mu|^{\frac13} \, .
\end{equation}
where $e(m) = \gamma(e, \dif m)= e^k\partial_k m$. The expression
above implies that all the metric concomitants can be written in
terms of $e$, and $\mu$ and its derivatives.

Now if we compute the Cotton-York tensor by using (\ref{nabla-e}),
we can solve equation (\ref{Ce}) and obtain $\chi$ up to an
undetermined function $\lambda$:
\begin{lemma} \label{lemma-chi-lambda}
For a type $D$ static vacuum solution, the gradient of the
potential, $\chi = \dif U$, takes the expression:
\begin{equation} \label{chi-lambda}
\chi = - \dif m + \lambda \, e  \,  ,  \qquad m \equiv \ln
|\mu|^{\frac13} \, ,
\end{equation}
where $e$ and $\mu$ are, respectively, the simple eigenvector and
the double eigenvalue of the Ricci tensor of the spatial metric
$\gamma$.
\end{lemma}

In order to determine the function $\lambda$, we put expression
(\ref{chi-lambda}) in the static vacuum equations (\ref{sve}) and,
if we make use of (\ref{nabla-e}), we obtain:
\begin{equation} \label{R-lambda}
R = -\nabla \dif m + \dif m \otimes \dif m + [e(\lambda) +
\lambda^2] e \otimes e -  \lambda \, e(m)  \gamma    \, .
\end{equation}

On the other hand, from the Ricci identities for the vector $e$ and
taking into account (\ref{nabla-e}) and (\ref{R-lambda}), we have:
\begin{equation} \label{lambda-e}
\lambda \, e(m) e = \dif [e(m)]   \, .
\end{equation}
When $e(m) \not=0$ the above equation enables us to determine
$\lambda e$
and from Lemma \ref{lemma-chi-lambda} we obtain an expression of $\chi$.\\[2mm]
{\bf Remark 1} All the type $D$ static vacuum metrics were obtained
by Elhers and Kundt \cite{ehlers-kundt} whom distinguished three
invariant classes: $A$, $B$ and $C$-metrics. The $B$-metrics are
those with a gradient of the Weyl eigenvalue laying on the
space-like principal two-plane (see also \cite{fs-Schw}, where the
intrinsic characterization of these three classes and of the
Schwarzschild solution was accomplished). Then, in terms of the
spatial metric $\gamma$ a $B$-metric is characterized by $e(m)=0$.

Consequently, so far our study has enabled us to obtain $\chi$ for
the $A$-metrics and $C$-metrics. In order to give its explicit
expression note that the modulus of the double eigenvalue may be
computed from $6 \mu^2 = \tr R^2$, and the projector on the simple
direction may be acquired from the equality $18 \mu^2 e \otimes e =
6 R^2 - s \gamma$. Moreover, (\ref{chi-lambda}) and (\ref{lambda-e})
show that actually we can obtain the potential $U$ or $V = e^U$ in
terms of the Ricci tensor. After all these considerations we can
state.

\begin{theorem} \label{theorem-AC}
Let $\gamma$ be a 3-dimensional metric, $R \equiv R(\gamma)$ its
Ricci tensor and $s \equiv  \tr R^2$, $t \equiv  \tr R^3$. The
necessary and sufficient conditions for $\gamma$ to be the spatial
metric of a $A$-metric or a $C$-metric are:
\begin{equation} \label{AC-1}
\tr R = 0  , \quad   6 t^2 = s^3 , \quad P(\dif s) \not= 0   , \quad
P  \equiv 6 R^2 - s \gamma   ,
\end{equation} \label{AC-2}
\begin{equation}
 R = \frac{1}{V} \nabla \dif V \, , \qquad V  \equiv
 [P(\dif s^{-2/3}, \dif s^{-2/3})]^\frac{1}{2} \, .
\end{equation}
\end{theorem}

The $A$-metrics are the vacuum type $D$ solutions with a gradient of
the Weyl eigenvalue laying on the time-like principal plane
\cite{fs-Schw}. Then, we can distinguish the $A$-metrics
(respectively, the $C$-metrics) by adding condition $P(\dif s)
\wedge \dif s=0$ (respectively, $P(\dif s) \wedge \dif s \not= 0$)
to the conditions of theorem \ref{theorem-AC}.

When $e(m)=0$ ($B$-metrics), making use of (\ref{ricci-D}) and
(\ref{nabla-e}), the $ee$-component of equation (\ref{R-lambda})
leads to:
\be \label{e-lambda}
e(\lambda) + \lambda^2 = - 2 \mu - (\dif m)^2 \, .
\ee

This last equation in $\lambda$ always admits solution. For every
solution $\lambda$ the vector $\chi$ given in (\ref{chi-lambda})
satisfies the static vacuum equations (\ref{sve}) which can now be
written substituting expression (\ref{e-lambda}) in equation
(\ref{R-lambda}). Then we obtain a condition solely involving
$\gamma$-concomitants which can be written in terms of the Ricci
tensor $R$.

\begin{theorem} \label{theorem-B}
Let $\gamma$ be a 3-dimensional metric, $R \equiv R(\gamma)$ its
Ricci tensor and $s \equiv  \tr R^2$, $t \equiv  \tr R^3$. The
necessary and sufficient conditions for $\gamma$ to be the spatial
metric of a $B$-metric are:
\begin{equation} \label{B-1}
\tr R = 0  , \quad   6 t^2 = s^3 ,
\quad P(\dif s) = 0   , \quad P  \equiv 6 R^2 - s \gamma   ,
\end{equation}
\begin{equation} \label{B-B}
R = \Omega^{-\frac13} \nabla \dif \Omega^{\frac13}  - 2[\Omega +
(\dif \Omega)^2] P \,  ,
\qquad \Omega \equiv -\frac{s}{18t}
\, .
\end{equation}
\end{theorem}
{\bf Remark 2} Once we have characterized the spatial metric
$\gamma$ of a static vacuum solution we can look for the potential
$U$ which completes the space-time metric $g$. Note that the
potential is defined up to an additive constant because only its
gradient appears in vacuum equations (\ref{sve}). One can take up
this constant by changing the static time $t$ with a constant
factor. Then, we can refer to the family of potentials that differ by
an additive constant as unique potential.

Theorems \ref{theorem-bartod} and \ref{theorem-AC} show that for
type I static vacuum solutions and for the $A$ and $C$-metrics the
potential $U$ is unique. Nevertheless, in the case of the
$B$-metrics the gradient of the potential depends on $\lambda$, a
solution of (\ref{e-lambda}). From (\ref{nabla-e}) a function
$\beta$ exists such that $e = \mu^{-\frac13} \dif \beta$. Moreover,
$\dif (\lambda e) = 0$ and, consequently, $\lambda = \mu^{\frac13}
f(\beta)$. Then, (\ref{e-lambda}) takes the expression:
\be \label{K}
f' + f^2 = K \, , \qquad K \equiv - \mu^{-1/3}[2 \mu + (\dif m)^2)]
\, .
\ee
From (\ref{B-1}) and (\ref{B-B}) we can show that the metric
concomitant $K$ is constant. Moreover it coincides with the
curvature of the two-dimensional metric of the time-like principal
plane of a $B$-metric \cite{fs-Schw}. For a $B_3$-metric one has
$K=0$ and equation (\ref{K}) admits solution $f=0$ and a
one-parametric family of solutions. For a $B_1$-metric ($K>0$) or a
$B_2$-metric ($K<0$) equation (\ref{K}) admits a one-parametric
family of solutions. In any case, one can obtain a canonical form
for the metric $\gamma$ and one shows that the associated space-time
metric $g$ is independent  of the potential $U$ obtained with the
different solutions $\lambda$.

The non uniqueness of the potential $U$ for a given spatial metric
$\gamma$ was underlined by Tod \cite{tod}. He found a family of
'spatial metrics which are static in many ways', and the resulting
space-time metrics $g$ are always of Petrov type $D$. He also claims
that all the $g$ associated with a $\gamma$ are diffeomorphic as we
have stated above. Our approach presented here also shows that
the spatial metrics studied by Tod are precisely those generating
the $B$-metrics.
\ \\[2mm]
{\bf Remark 3} In this paper we have accomplished the intrinsic
characterization of the 3-metrics $\gamma$ which are the spatial
metric of a static vacuum solution $g$. A different problem, albeit
with a similar statement is to intrinsically characterize the
space-time metric $g$ itself. We have solved this problem for type
$D$ static vacuum solutions \cite{fs-Schw}, and in \cite{fms} we
have given an intrinsic characterization of the type I static
metrics.
Theorems \ref{theorem-bartod}, \ref{theorem-AC} and \ref{theorem-B}
can be summarized in the following algorithm presented below
as a flow chart. Horizontal arrows pointing the symbol $[\bullet]$
indicate that $\gamma$ is not the spatial metric of a static vacuum
solution.

\vspace*{1.7cm} \setlength{\unitlength}{0.85cm} {\footnotesize
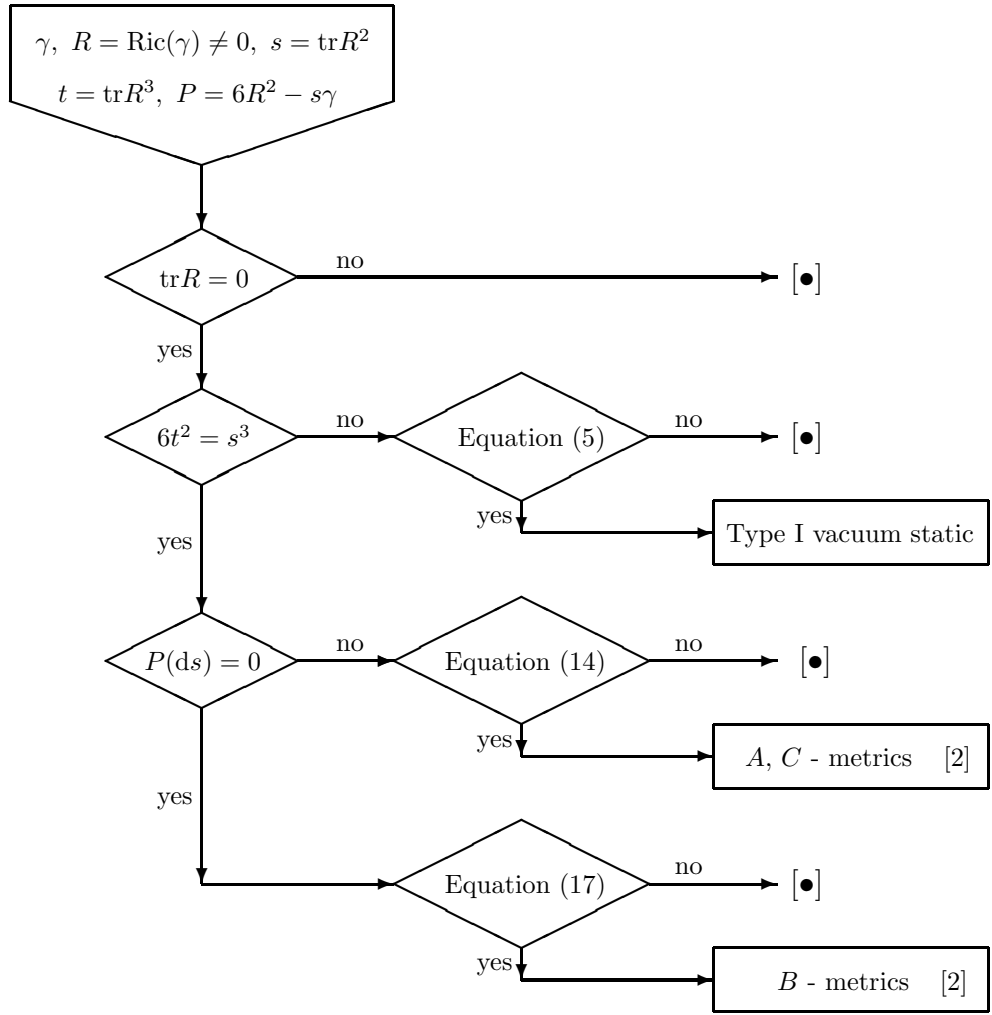
\noindent
\begin{picture}(0,18)
\thicklines

\put(3,17){\line(3,1){3}}

\put(0,18){\line(3,-1){3}}

\put(0,18 ){\line(0,1){1.5}}\put(6,19.5){\line(-1,0){6}}

\put(6,19.5){\line(0,-1){1.5}}

\put(0.4,18.8){$ \gamma, \ R={\rm Ric}(\gamma) \neq 0 , \ s= {\rm
tr} R^2 $}

\put(0.75,18){$t= {\rm tr} R^3 , \  P= 6 R^2 - s \gamma$}

 \put(3,17){\vector(0,-1){1}}

\put(1.5,15.25){\line(2,1){1.5}}\put(1.5,15.25){\line(2,-1){1.5}}
\put(4.5,15.25){\line(-2,1){1.5}} \put(4.5,15.25){\line(-2,-1){1.5}}

\put(2.2,15.1){ ${\rm tr} R =0 $}

\put(3,14.5){\vector(0,-1){1}}

\put(1.5,12.75){\line(2,1){1.5}}\put(1.5,12.75){\line(2,-1){1.5}}
\put(4.5,12.75){\line(-2,1){1.5}} \put(4.5,12.75){\line(-2,-1){1.5}}

\put(3,12){\vector(0,-1){2}}

\put(4.5,15.25){\vector(1,0){7.5}}

\put(4.5,12.75){\vector(1,0){1.5}} \put(10,12.75){\vector(1,0){2}}

\put(8,11.75){\vector(0,-1){0.5}} \put(8,11.25){\vector(1,0){3}}

\put(6,12.75){\line(2,1){2}}\put(6,12.75){\line(2,-1){2}}
\put(10,12.75){\line(-2,1){2}} \put(10,12.75){\line(-2,-1){2}}

\put(1.5,9.25){\line(2,1){1.5}}\put(1.5,9.25){\line(2,-1){1.5}}
\put(4.5,9.25){\line(-2,1){1.5}} \put(4.5,9.25){\line(-2,-1){1.5}}

\put(6,9.25){\line(2,1){2}}\put(6,9.25){\line(2,-1){2}}
\put(10,9.25){\line(-2,1){2}} \put(10,9.25){\line(-2,-1){2}}

\put(8,8.25){\vector(0,-1){0.5}} \put(8,7.75){\vector(1,0){3}}

\put(4.5,9.25){\vector(1,0){1.5}}

\put(3,5.75){\vector(1,0){3}} \put(3,8.5){\vector(0,-1){2.75}}

\put(6,5.75){\line(2,1){2}}\put(6,5.75){\line(2,-1){2}}
\put(10,5.75){\line(-2,1){2}} \put(10,5.75){\line(-2,-1){2}}

\put(8,8.25){\vector(0,-1){0.5}}

\put(4.5,9.25){\vector(1,0){1.5}} \put(10,9.25){\vector(1,0){2}}

\put(8,4.75){\vector(0,-1){0.5}} \put(8,4.25){\vector(1,0){3}}

\put(10,5.75){\vector(1,0){2}}

\put(11,8.25){\line(1,0){4.3}} \put(11,7.25){\line(0,1){1}}
\put(15.3,7.25){\line(-1,0){4.3}} \put(15.3,8.25){\line(0,-1){1}}

\put(11,4.75){\line(1,0){4.3}} \put(11,3.75){\line(0,1){1}}
\put(15.3,3.75){\line(-1,0){4.3}} \put(15.3,4.75){\line(0,-1){1}}

\put(11,11.75){\line(1,0){4.3}} \put(11,10.75){\line(0,1){1}}
\put(15.3,10.75){\line(-1,0){4.3}} \put(15.3,11.75){\line(0,-1){1}}

\put(2.3,12.6){$6 t^2 = s^3$} \put(2.1,9.1){$P({\rm d} s) =0$}

\put(11.2,11.1){Type I vacuum static }

\put(11.5,7.6){$A$, $C$ - metrics $\ \,  $ \cite{ehlers-kundt}}

\put(12,4.1){$B$ - metrics $\ \, $  \cite{ehlers-kundt}}

\put(2.3,14){\footnotesize yes} \put(2.3,7){\footnotesize yes}
\put(2.3,11){\footnotesize yes}

\put(7.3,7.9){\footnotesize yes}

\put(7.3,4.4){\footnotesize yes} \put(7.3,11.4){\footnotesize yes}

\put(5.1,15.4){\footnotesize no}

\put(5.1,12.9){\footnotesize no} \put(10.4,12.9){\footnotesize no}

\put(5.1,9.4){\footnotesize no} \put(10.4,9.4){\footnotesize no}

\put(10.4,5.9){\footnotesize no}

\put(12.2,5.6){\normalsize $[\bullet ] $ }

\put(12.2,9.1){ {\normalsize $ [\bullet]$ }}

\put(12.2,12.6){\normalsize $[\bullet]$   }

\put(12.2,15.1){\normalsize $[\bullet]$   }

\put(7 ,12.6){\footnotesize Equation (5)}

\put(6.8 ,9.1){\footnotesize Equation (14)}

\put(6.8 ,5.6){\footnotesize Equation (17)}
\end{picture}

}
\vspace*{-3.5cm}

\ack This work has been supported by the Spanish ministries of
``Ciencia e Innovaci\'on" and ``Econom\'{\i}a y Competitividad",
MICINN-FEDER projects FIS2009-07705 and FIS2012-33582.

\end{document}